\documentclass[superscriptaddress,aps,amssymb,amsfonts,twocolumn]{revtex4-1}
\usepackage{times,xspace}
\usepackage{amsbsy,amssymb,amsmath,bm,appendix}
\usepackage{graphicx,color,epsfig,rotate}
\usepackage{fancyhdr}

\def\one{{\mathchoice {\rm 1\mskip-4mu l} {\rm 1\mskip-4mu l} {\rm
1\mskip-4.5mu l} {\rm 1\mskip-5mu l}}}
\def\bbbc{{\mathchoice {\setbox0=\hbox{$\displaystyle\rm C$}\hbox{\hbox
to0pt{\kern0.4\wd0\vrule height0.9\ht0\hss}\box0}}
{\setbox0=\hbox{$\textstyle\rm C$}\hbox{\hbox
to0pt{\kern0.4\wd0\vrule height0.9\ht0\hss}\box0}}
{\setbox0=\hbox{$\scriptstyle\rm C$}\hbox{\hbox
to0pt{\kern0.4\wd0\vrule height0.9\ht0\hss}\box0}}
{\setbox0=\hbox{$\scriptscriptstyle\rm C$}\hbox{\hbox
to0pt{\kern0.4\wd0\vrule height0.9\ht0\hss}\box0}}}}

\newcommand{\ket}[1]{|{#1}\rangle}
\newcommand{\bra}[1]{\langle{#1}|}

\newcommand{\ignore}[1]{}
\newcommand{\mComment}[1]{}
\newcommand{\gComment}[1]{}
\newcommand{\jComment}[1]{}
\newcommand{\rComment}[1]{}
\newcommand{\lComment}[1]{}

\def\cV{{\cal V}}
\def\cH{{\cal H}}
\def\cO{{\cal O}}

\renewcommand{\jComment}[1]{\textcolor{green}{Cristian: #1}}

\begin{document}

\title{An exact real-space renormalization method and applications}

\author{Adrian E. Feiguin}
\email{a.feiguin@neu.edu}
\affiliation{Department of Physics, Northeastern University, Boston, MA 02115, USA.}

\author{Rolando D. Somma}
\affiliation{Theory Division, Los Alamos National Laboratory, Los Alamos, NM 87545, USA.}
\email{somma@lanl.gov}

\author{Cristian D. Batista}
\affiliation{Theory Division, Los Alamos National Laboratory, Los Alamos, NM 87545, USA.}

\email{cdb@lanl.gov}

\date{\today}

\begin{abstract}
We present a numerical method based on real-space renormalization
that outputs the exact ground space of ``frustration-free" Hamiltonians.
The complexity of our method is polynomial in the degeneracy 
of the ground spaces of the Hamiltonians involved in the renormalization steps.
We apply the method to obtain the full ground spaces of two spin systems.
 The first system is
a spin-1/2 Heisenberg model with  four-spin cyclic-exchange interactions defined on a square lattice.
In this case,  we  study finite lattices of up to 160 spins and find a triplet ground state that differs from the singlet ground states obtained in C.D. Batista and S. Trugman, Phys. Rev. Lett. {\bf 93}, 217202 (2004). We characterize such 
a triplet state as consisting  of a triplon that propagates in a  background of fluctuating singlet dimers.  
The second system is a family of spin-1/2 Heisenberg chains with uniaxial exchange anisotropy and  next-nearest neighbor interactions. In this case, the method finds a ground-space degeneracy that scales quadratically with the system size and outputs the full ground space efficiently. Our method can substantially outperform methods based on exact diagonalization and is more  efficient than other renormalization methods  when the ground-space degeneracy is large.


\end{abstract}

\maketitle

\section{Introduction}
Renormalization methods are powerful tools for studying the long-wavelength properties of physical systems by  a systematic elimination of high-energy degrees of freedom. The first numerical  renormalization group (NRG) method was developed by Wilson \cite{Wilson75,BullaRMP08}  to solve the Kondo problem, an important problem in physics that involves the interaction of a magnetic impurity with a conduction band \cite{Kondo64}. The more recent density-matrix renormalization group   (DMRG) method was successfully applied to a large class of one-dimensional   ($D=1$) quantum systems \cite{White92} and a few $D=2$ systems \cite{steve kagome, peps matthias, uli kagome, balents j1j2}. Recent advances in quantum information theory also led to renormalization and variational methods, including PEPS~\cite{peps2d,Sab08}, MERA \cite{mera, mera2d}, and tensor renormalization  \cite{trg, levin, tao, tao2}. 
The problem with  known  renormalization  methods is that they suffer from important limitations  when
studying systems in space dimension $D \geq 2$ or with a large number of ground states. 
Our goal is  to construct a renormalization method that can be applied to such systems 
when the Hamiltonians under consideration satisfy a  ``frustration-free" (FF) property. 

The term ``frustration free'' was  first coined by the quantum-information community~\cite{Perez08,
Bravyi2009,Somma2011} to denote a class of Hamiltonians  $H = \sum_{k=1}^p \pi_{v_k} $
whose ground states are also ground states of each local term $\pi_{v_k}$.  $v_k$ refers
to a finite set of degrees of freedom, e.g., a unit or a finite subsystem.
While describing such Hamiltonians as FF is adequate from a viewpoint that we discuss below,  
the term FF can be confusing
if we adopt a more traditional convention of identifying frustration with competing interactions.
For example, the 
 triangular lattice Ising model with antiferromagnetic (AFM) exchange ($J>0$) is the paradigmatic example of a frustrated Hamiltonian. However, this model is FF according to the previous definition. 
We let $v_k$ be the three spins  $\sigma^k_j =\{-1,1\}$ 
in the $k$th triangle, $j=\{1,2,3\}$, and define $\pi_{v_k}= (J/2) [(\sum_{j} \sigma^k_j )^2-1]$.
The Hamiltonian is FF because any ground state $\ket \psi$ of  $H$ satisfies $ \pi_{v_k} \ket \psi =0$, i.e., $\ket \psi$ is also a ground state of each $\pi_{v_k}$. 
Nevertheless, $\ket \psi$ does not {\em minimize} each of the  bond  Hamiltonians $ J \sigma^k_j \sigma^k_{j'} $,  reason why the model  is considered to be frustrated
according to the traditional convention.

The previous discussion  implies that
the concept of frustration is  relative to a particular decomposition of  $H$. The traditional interpretation of frustration assumes a  decomposition of $H$
dictated by the physical nature of the interactions.
However, while $H$ may be frustrated with respect to one decomposition,
it may still be FF  because 
the competition between interactions on different units disappears when we consider a different decomposition (e.g., triangles instead of bonds in the Ising example).
Remarkably,
FF Hamiltonians are ubiquitous in condensed matter and quantum information
theory. They include Ising models, the AKLT model~\cite{AKLT}, parent Hamiltonians of PEPS~\cite{peps2d}, and Hamiltonians
that can simulate quantum circuits~\cite{Aharonov2004}.
Several other  frustrated magnets also correspond to FF Hamiltonians~\cite{Batista12}.
Ground states of FF Hamiltonians contain all the characteristics of highly frustrated physical systems: large ground state degeneracy~\cite{Wannier50}, coexistence of different phases, and  exotic orderings.

In this manuscript, we introduce an exact real-space renormalization method (ERM) that 
obtains the full ground-space of FF Hamiltonians. 
The output of the ERM is a sequence of tensors
whose contraction allows us to compute expectation values of observables
and amplitudes of the ground states (Sec.~\ref{sec:RM}).
The computational cost of our method (i.e., the cost of the tensor contraction)
is polynomial in the ground-space degeneracy
of the FF Hamiltonians involved in the renormalization steps.
If such a degeneracy increases  polynomially with the system size,
the ERM is efficient. Otherwise, for exponentially large degeneracies,
the ERM is inefficient but can  substantially outperform other numerical techniques
for this problem.

To illustrate the potential of our  method, we apply it to two 
FF spin systems that have largely degenerate ground states. The first system  is a spin-1/2 Heisenberg model with  four-spin cyclic-exchange interactions that is defined on a square lattice (Sec. \ref{sec:2Dmagnet}). The ground state degeneracy is  exponential in the linear size, $L$, of the lattice. Such a degeneracy is  much smaller than the Hilbert space dimension $2^{L^2}$, so the ERM outperforms exact diagonalization in this case.
Besides the singlet ground states that were in identified in Ref.~\cite{Batista04}, we find a triplet ground state that consists of a triplon that propagates in a background of fluctuating singlet dimers. The propagation of this triplon leads to an incipient long range  AFM ordering, which indicates that the triplet ground state describes an AFM quantum critical point.  Such a triplet ground state also exists in rectangular spin lattices of size $ L_x \times L_y$,
$L_x \leq L_y$,  with  periodic boundary conditions. In particular, Lajk\'{o},  Sindzingre, and  Penc, gave an analytical expression of this state for the case of three-leg tubes ($L_x=3$) \cite{Lajko12}. 

The second of system consists of the family of  spin-1/2 Heisenberg chains
with uniaxial exchange anisotropy and nearest and next-nearest-neighbor exchange interactions
(Sec.~\ref{sec:1DHeisenberg}). The Hamiltonians in this family are also FF. An analytical and
 closed form representation for the ground states of this family, in terms of anyonic operators, was given in Ref.~\cite{Batista12}.
However, such a representation may not be useful for computing  some expectation values of spin-spin correlations efficiently. Those expectation values can be efficiently computed with the ERM.

We note that quantum Monte Carlo methods, that are not based on renormalization,  cannot be applied to most FF Hamiltonians because of the infamous sign problem.




\section{The renormalization method}
\label{sec:RM}
\begin{figure}[ht]
  \centering
  \includegraphics[width=6.5cm]{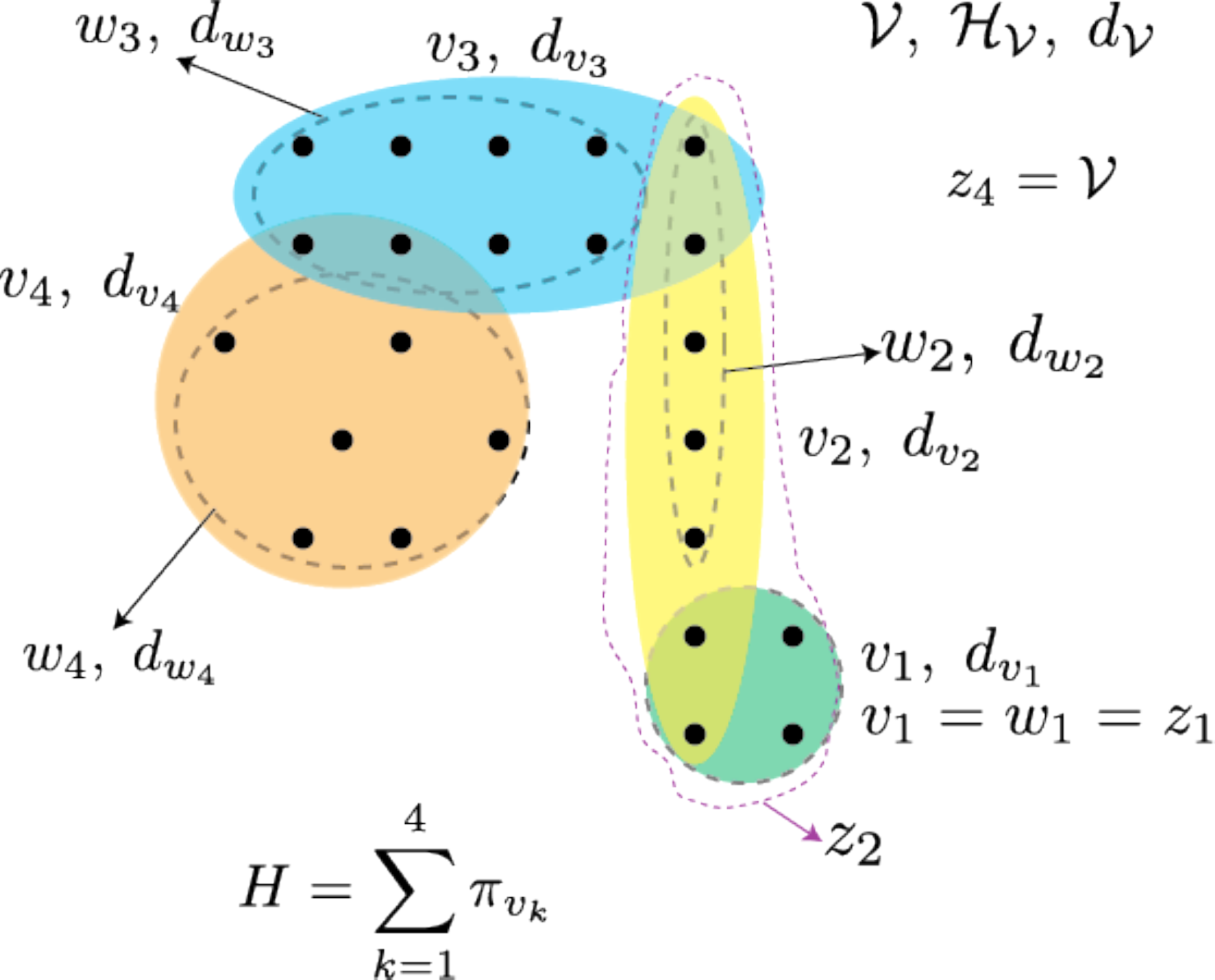}
  \caption{Spin system. Black dots denote spins located
  at the vertices $\cV$ of the lattice. Each term $\pi_{v_k}$ in the Hamiltonian
  acts nontrivially in $v_k$ only. The sets  $w_l$ and $z_l$ are defined in the text.
}
  \label{fig:RGFig1}
\end{figure}

We start by providing a brief description of the ERM (technical details are provided in Appendix~\ref{app:RM}).
For simplicity, we consider a Hamiltonian $H$ acting
on a system of $L$ spins located on the vertices $\cal V$ of a lattice~\cite{NOTE2}. 
The Hilbert space of the system is $\cH_{\cV}$ and its dimension is $d_{\cV}$.
The magnitude of the spins and space dimensionality of the lattice are arbitrary;
we refer to the spin 
system of Fig.~\ref{fig:RGFig1} for illustration purposes. We let $v_k$ be a subset of spins
of $\cV$, $\cH_{k}$ the associated Hilbert space, and $d_{v_k}$ its dimension.
Here, $ k=1,2,\ldots,p $ and the subsets $v_k$ are known. 
For $l=1,2,\ldots,p$ we also define the sets of spins $w_{l}=v_{l} \setminus z_{l-1}$, with
$z_l = \cup_{k=1}^l v_k =  \cup_{k=1}^l w_k$ and $z_0 = \{ \emptyset \}$.
  $A \setminus B$ is the relative complement of $A$ in $B$ so that $w_{l}$ are those spins
that belong to $v_{l}$ but do not belong to any other $v_k$ with $1\le k \le l-1$. We assume
$w_l \ne \{ \emptyset \}$ and note   that $v_1 = w_1 =z_1$.

 For a set $x$ of spins in $\cV$, we use $i^x$ for the spin variable in some standard basis
~\cite{NOTE1}.
Also, $d_x$ denotes the dimension of $\cH_x$, the Hilbert space associated with  $x$. 
It follows that  $\{ \ket {i^{x}} \}_{1 \le i^x \le d_x}$ is an orthonormal basis for the 
spins in $x$; hereafter referred to as the computational basis and $\ket{i^x}$ is a basis state.
In some cases, we do not distinguish between basis states  or (column) vectors:
$\ket {i^{x}}$ can also denote a vector with component equal to 1 in position
$ i^x$ and zeroes elsewhere. The number of components of 
$\ket {i^{x}}$ is the number of values that $i^x$ can take
($\le d_x$), which is given in each case.

\begin{figure}[ht]
  \centering
  \includegraphics[width=6.5cm]{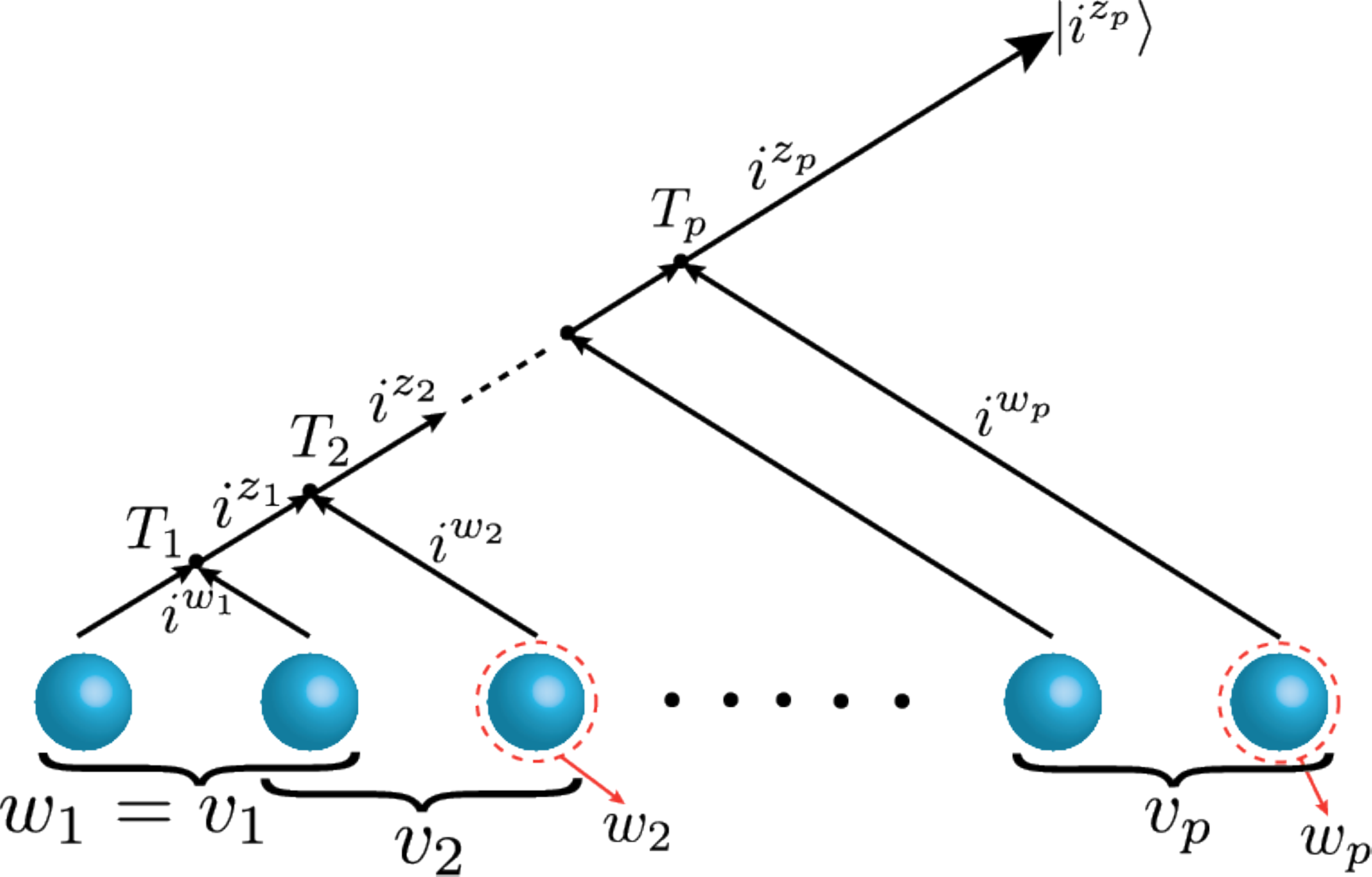}
  \caption{Representation of the tree-tensor network contraction
    for
  a  system with $p$ spins (blue circles). In this example, the sets $v_l$
  refer to pairs of nearest-neighbor spins and each $w_l$
  is a single spin. The tensors $T_l$
  live in the vertices of the tree and the contraction indices are
  in the edges. An arrow means a sum over the corresponding index.
  The contraction $T_p \bullet \ldots \bullet T_1$ maps ground states of $H$ 
  into states $\ket{i^{z_p}}$.  $T^\dagger_1 \bullet \cdots \bullet T^\dagger_p$ is the inverse transformation
  that gives all the ground states of $H$ from $\ket{i^{z_p}}$.
  }
  \label{fig:RGFig2}
\end{figure}

The Hamiltonian is represented as
\begin{align}
H = \sum_{k=1}^p \pi_{v_k} \; ,
\end{align}
where each $\pi_{v_k}$ is a Hermitian operator acting nontrivially on spins in $v_k$ only.
We assume $\pi_{v_k} \ge 0$.
If any ground state $\ket \psi$  of $H$
satisfies
\begin{align}
\label{eq:FFproperty}
 H \ket \psi = \pi_{v_1} \ket \psi = \ldots =\pi_{v_p} \ket \psi   =0 \; ,
\end{align}
then $H$ is said to be FF. The standard definition of FF also assumes
that each $\pi_{v_k}$ is local. Here, we can relax such an assumption
without incurring in large computational overheads 
as long as $\pi_{v_k}$ is a bounded sum of product operators (see Appendix~\ref{app:RM}).
When $H$ is FF, our renormalization method is {\em exact}
 and outputs  the ground states of $H$ as 
  \begin{align}
 \label{eq:groundstates}
 \ket{\psi_{ i^{\cV}}^p} =   T^{\dagger}_1  \bullet \ldots \bullet  T^{\dagger}_p\ket{ i^{\cV}} \; .
 \end{align}
  $T^\dagger_1,T^\dagger_2,\ldots ,T^\dagger_p$ are isometries,
 $1 \le i^{\cV} \le g$, and $g$ is the ground-space dimension.
 The symbol $\bullet$ refers to 
 a tensor contraction that
 involves a sum over repeated indices. Such a contraction
 regards a tree-tensor network (see Fig.~\ref{fig:RGFig2});
 Tree-tensor networks are ubiquitous  in renormalization methods~\cite{Vid2007,
 EV2012}.
 Whether $H$ is FF or not  is also an output of the ERM.

 The ERM performs $p$ steps. Each step can be defined recursively as follows. 
 For $1 \le l \le p$, we let $g_l \ge 0$ be the ground-space degeneracy
 of $H_l = \sum_{k=1}^l \pi_{v_k}$ ($g_0=1$). The ground states
 of $H_l$ are $\ket{\psi^l_{i^{z_l}}}$, $1 \le i^{z_l} \le g_l$.
  $H_p=H$. In the first step, the ERM diagonalizes $\gamma_1$,
  the $d_{w_1} \times d_{w_1} $ matrix representation
  of $\pi_{v_1}$ in the computational basis. It obtains
  $g_1$ and $\{   \ket{\psi^1_{i^{z_1}}}   \}_{1 \le i^{z_1} \le g_1}$, and continues
 only if $g_1>0$.
  In the $l$-th step, $l \ge 2$, the ERM computes $\gamma_l$. This is
  a $h_l \times h_l$ matrix representation of $\pi_{v_l}$
  in the basis $\{ \ket{\psi^{l-1}_{i^{z_{l-1}}}, i^{w_l}} \}$, with 
  $1 \le i^{z_{l-1}} \le g_{l-1}$, $1 \le i^{w_l} \le d_{w_l}$, and $h_l = g_{l-1} d_{w_l}$.
  Then ERM applies exact diagonalization to $\gamma_l$,
  obtains a basis $\{ \ket{\psi^{l}_{i^{z_{l}}} }\}$  for the zero eigenvalue, and  
  computes the multiplicity $g_l$ of the new ground space. 
 The ERM continues only if $g_l >0$. The isometries
  in Eq.~\eqref{eq:groundstates} are
  \begin{align}
  T^\dagger_l = \sum_{i^{z_l}=1}^{g_l} \ket{\psi^l_{i^{z_l}}} \bra {i^{z_l}} \; .
  \end{align}

 In  Appendix~\ref{app:RM}
 we show that, if $N_T$ is the number of elementary
 operations to output the isometries
 $T^\dagger_1, \ldots, T^\dagger_p$, then 
 \begin{align}
 N_T \propto \sum_{k=1}^p EV( h_k ) + (p-k) (h_k g_{k})^2 \; .
 \end{align}
  $EV(d)$ is the cost of the exact diagonalization of a $d \times d$ matrix and $g_0=1$.  
  Also, if $M_T$ is the memory cost associated with the number
  of variables kept  during the implementation of the ERM, 
  \begin{align}
  M_T \propto \sum_{k=1}^p h_k g_k \; .
  \end{align}
  Thus, the efficiency of the ERM strongly depends on $g_k$ 
  and the method becomes efficient when $g_k \in \cO[{\rm poly} (p)]$.
The cost of evaluating expectation values
of observables in any ground state of $H$ is also important
and can be easily derived from the analysis given in Appendix~\ref{app:RM}.

 A related method for solving some spin-1/2 systems, which is based
on the techniques developed in Ref.~\cite{Bravyi2011}, can be found in Ref.~\cite{BOO2010}.
Also, an exact renormalization method for quantum spin chains was proposed in
Ref.~\cite{HS10}. 
Our main contribution with respect to that of Refs.~\cite{BOO2010,HS10} is that we consider 
  Hamiltonians with arbitrary interactions, in any space dimension, and provide a  real-space renormalization algorithm that is
exact if the ground space satisfies some properties that can be verified by the ERM. 
 In addition, the way that the ERM contracts tensors is different from the usual contraction 
of a binary-tree like tensor network.  
The main reason behind this difference
is the minimization of $M_T$ when $g_l$ grows monotonically with $l$.


\section{Application to a $D=2$ magnet}
\label{sec:2Dmagnet}
\subsection{Model Hamiltonian}

\begin{figure*}[t!]
\includegraphics[width=0.7\textwidth,angle=0]{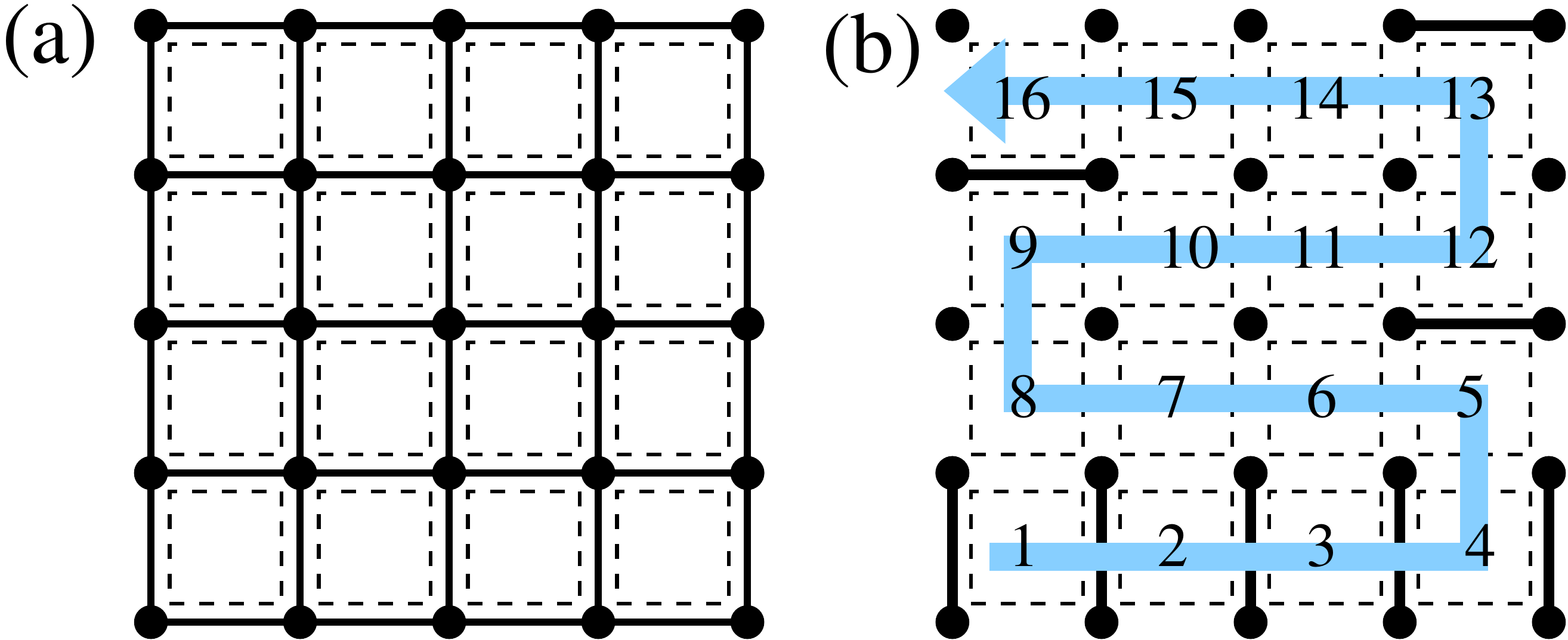}
\caption{(Color online) (a) Square lattice and the projectors $\cal{P}^{\boldsymbol{\alpha}}$ acting on the square plaquettes (dashed lines). Each $\cal{P}^{\boldsymbol{\alpha}}$ projects the spin  states of the plaquette onto the subspace with  spin $2$. (b) ``Snake'' path followed by the ERM, similar to the DMRG warmup scheme. At each step one adds the spins necessary to complete a projector on a new plaquette. This  growing process requires adding one or two spins (thick bold lines) per step. }
\label{fig:lattice}
\end{figure*}

We apply the ERM  to a spin-$1/2$ model that satisfies the frustration-free property
as defined in Sec.~\ref{sec:RM}. The FF Hamiltonian is defined on a  square lattice and reads \cite{Batista04}:
\begin{eqnarray}
H = \frac{3}{2}\sum_{\boldsymbol {\alpha}} {\cal P}^{\boldsymbol {\alpha}} \; .
\label{Hamil2}
\end{eqnarray}
Each operator ${\cal P}^{\boldsymbol {\alpha}}$ projects
the total spin state of a square plaquette $\boldsymbol {\alpha}$ onto the subspace with   spin $2$, as illustrated in Fig.\ref{fig:lattice}(a). This model has a number of exact valence-bond ordered ground states that increases exponentially in the linear dimension of the square lattice~\cite{Batista04}. 
The model of Eq.~\eqref{Hamil2} corresponds to a particular  regime of parameters of a more general model with frustration and ring exchange:
\begin{eqnarray}
H' &=& J_1 \sum_{\langle  {\bf r} , {\bf r}'  \rangle} {\bf s}_{\bf r} \cdot {\bf s}_{{\bf r}'}
+J_2 \sum_{\langle \langle  {\bf r} , {\bf r}'  \rangle \rangle} {\bf s}_{\bf r} \cdot {\bf s}_{{\bf r}'}
\nonumber \\
&+&  K \sum_{\boldsymbol {\alpha}} (P^{\boldsymbol {\alpha}}_{ij} P^{\boldsymbol {\alpha}}_{kl}
+P^{\boldsymbol {\alpha}}_{jk} P^{\boldsymbol {\alpha}}_{il} 
+P^{\boldsymbol {\alpha}}_{ik} P^{\boldsymbol {\alpha}}_{jl}) \; .
\label{Hamil}
\end{eqnarray}
Here,  $\langle  {\bf r} , {\bf r}'  \rangle$ and
$\langle \langle  {\bf r} , {\bf r}' \rangle \rangle$ denote nearest neighbors
and next nearest neighbors, respectively, and $i$, $j$, $k$, and $l$ label the four spins of a 
 plaquette in cyclic order. ${\bf s_r}=(s^x_{\bf r},s^y_{\bf r},s^z_{\bf r})$ is the spin operator of the spin at the ${\bf r}$th position.
Equation \eqref{Hamil2} reduces to Eq. \eqref{Hamil}, up to an irrelevant constant, if $J_1=1$, $J_2=1/2$, 
and $K=1/8$. This case corresponds to the maximally frustrated point as a function of $J_2/J_1$.

\subsection{Algorithm}

The implementation of the ERM is related to Wilson's NRG \cite{Wilson75,BullaRMP08} and to the warmup stage of the conventional DMRG \cite{White92}. Thus, it is simple to adapt an existing NRG or DMRG code 
for this case by making minor modifications. At each renormalization step, the ERM {\em grows} the system 
by adding one plaquette.  Then, the ERM diagonalizes the renormalized Hamiltonian and only keeps
the ground states (if the lowest eigenvalue is zero). In a DMRG ``language'', this corresponds to working with two blocks instead of four.

The order in which the plaquettes are added can be arbitrary.
However, the ground space degeneracy, $g_l$,  
depends dramatically on the path  followed to grow  the  lattice. 
 The ``snake'' path shown in  Fig. \ref{fig:lattice}(b) is the approach that turned out to be more efficient. 
The number of spins increases by one  or by two  at each step [see for instance steps 5 and 6 in Fig. \ref{fig:lattice}(b)].
It is important to introduce  Hamiltonian terms corresponding to a plaquette at a time (the projectors $\cal{P}^{\boldsymbol{\alpha}}$) and to avoid including terms belonging to neighboring plaquettes. For instance, a nearest neighbor exchange term in Eq.~\eqref{Hamil} is shared by two neighboring plaquettes. Such a term should be split accordingly to assure that only one projector is added per step. 

We applied the ERM  to rectangular lattices of sizes $L_x \times L_y$, $L_x \le L_y$, and periodic boundary conditions. 
The ground space degeneracy $g$ is proportional to $2^{L_x}$ in this case, i.e. exponential in the linear dimension.
If we follow a path like the ``snake'' in Fig. \ref{fig:lattice}(b), $g_l$ increases as we increase $L_x$ but it does not change
substantially  when we increase $L_y$.
This property allows us to study remarkably large system sizes by keeping
$L_x$  constant and by increasing the number of plaquettes in
the $y$ direction to relatively large values of $L_y$.
 Every time we close a boundary along the $y$ direction, the degeneracy $g_l$
drops substantially.


Unlike DMRG, the ERM needs to fully diagonalize the renormalized Hamiltonian at each step. 
This implies diagonalizing a $h_l \times h_l$ matrix in the $l$th step, with $h_l =2 g_{l-1}$ or $h_l = 4 g_{l-1}$,
depending on the number of spins added at that step (one or two, respectively).
The maximum value of $h_l$ depends on the linear dimensions of the system, and could reach several thousands
for lattices of hundreds of spins.  To be able to study relatively large systems, we can use symmetries --U(1)/abelian quantum numbers in our case-- to store the  Hamiltonians and other operators in block form. However, 
to ensure that we keep all the ground states,  we do not restrict the values of these quantum numbers.

\begin{figure}[ht]
 \centering
 \includegraphics[width=0.35\textwidth,angle=-90]{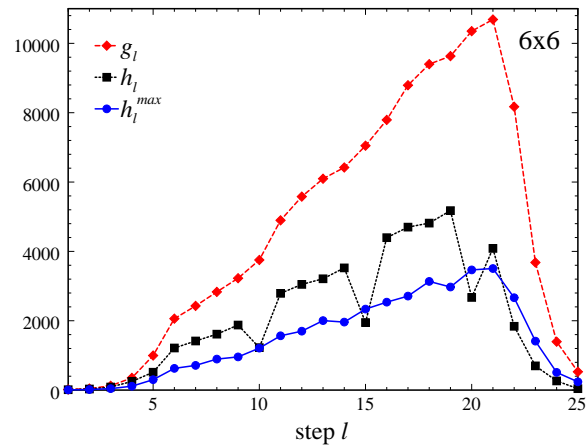}
 \vspace*{-0.2cm}
 \caption{Ground space degeneracy ($g_l$), order of the renormalized Hamiltonian ($h_l$),
  and order of the largest block in the renormalized Hamiltonian ($h_l^{\max}$)  per step
  of the ERM, when
  symmetries are considered.}
  \label{fig:matrix}
\end{figure}

In Fig. \ref{fig:matrix} we plot  the ground-space degeneracy, $g_l$, and order of the renormalized Hamiltonian,  $h_l$, for a $6 \times 6$ lattice ($L_x=L_y=6$). We also plot the  order of the largest block
of the renormalized Hamiltonian when symmetries  are considered, $h_l^{\max}$,
which determines the dominating cost of the method.
The oscillation in  $g_l$ has a periodicity corresponding to the linear dimension of the lattice,
showing the reduction in the ground space degeneracy every time a lattice boundary is closed.

 Because we are interested in the computation of correlation functions
of arbitrary ground states, at each step we need to store all the matrices of the operators
involved in such correlations. We note that correlators of the form $\langle AB \rangle$
cannot be computed by storing the operators $A$ and $B$ independently. In general, it is necessary  to store
the matrices for the product $AB$ because the product of two projected operators is not equal to  the
projection of the product.


 The ground states of $H$ can be obtained for $8\times8$ and larger lattices by implementing the ERM on more powerful existing computers. The requirement of storing all the operators associated with the correlations described in the next section and those needed for computing the Hamiltonian terms,  is the main limiting factor for increasing the 
lattice size. 

\subsection{Results}

\begin{figure}[ht]
  \centering
  \includegraphics[height=8.5cm,angle=0]{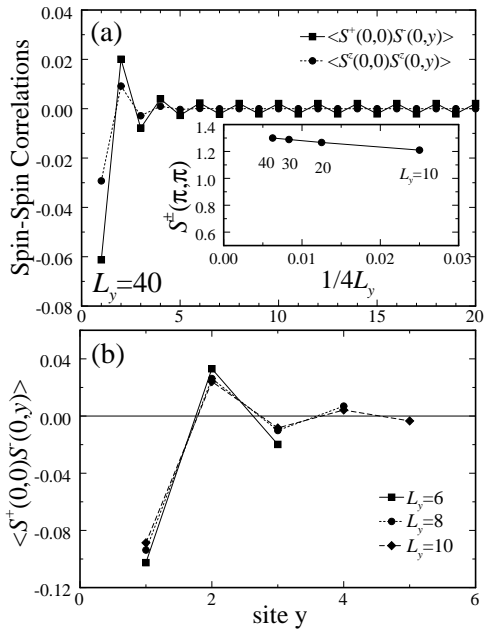}
  \caption{ Two-point correlators $\langle s^z_{\bf 0} s^z_{\bf r}  \rangle$ and 
  $\langle  s^+_{\bf 0} s^-_{\bf r}  \rangle$ as a function of distance along the $y$-direction.
  Here, ${\bf 0}$ is a spin of reference and ${\bf r}=(0,y)$. The figure shows the correlations 
 for a) $4 \times L_y$ and b) $6 \times L_y$ lattices.
Inset of a): scaling  of the structure factor, $S^{\pm}({\bf k})$, evaluated at the AFM wave vector ${\bf k}=(\pi,\pi)$.}
  \label{fig:4}
\end{figure}

In addition to the large set of exact $S=0$ ground states exhibiting valence bond ordering~\cite{Batista04}, the  results output by the ERM show a 
 triplet  ground state, with $S=1$, $S^z=\pm1,0$. ($S$ is the total spin of the lattice and $S^z$ the $z$th component of the total spin.) Such a state was also identified by exact diagonalization of small square clusters and by an analytical solution of the model on $3\times L_y$ tubes~\cite{note1,Lajko12}. That the $S=1$ state exists
 in larger systems was unexpected and illustrates  the importance of methods that obtain the full ground space.
 The existence of a triplet ground state is compatible with a critical scenario in which $H$ has a gapless spectrum of $S=1$ spin excitations. In this case,  an external magnetic field $B>0$ would induce AFM ordering of the spin components that are orthogonal to the  field's direction.
This scenario is confirmed by the spin-spin correlators obtained with our ERM for the $S= S^z=1$ ground state on $4 \times L_y$ and $6 \times L_y$ finite lattices. Figure~\ref{fig:4} shows the two-point correlators $\langle { s}^z_{\bf 0} {s}^z_{\bf r}  \rangle$ and 
$\langle  { s}^+_{\bf 0} { s}^-_{\bf r}  \rangle$ as a 
 function of the distance along the $y$-direction. ${\bf 0}=(0,0)$ denotes a reference spin (the origin), ${\bf r}=(0,y)$, and
$s^\pm_{\bf r} = s^x_{\bf r} \pm i s^y_{\bf r}$. While there is a clear long-range AFM tail for $\langle  { s}^+_{\bf 0} {s}^-_{\bf r}  \rangle$, the $\langle { s}^z_{\bf 0} { s}^z_{\bf r}  \rangle$ correlator decays exponentially in $r$. It is important to note that the magnitude of the local staggered $xy$-magnetization is of order $1/\sqrt{L_x L_y}$, which is the expected behavior for the condensation of a single triplon. This property is revealed by the scaling behavior of the structure factor,
\begin{eqnarray}
S^{\pm}({\bf k}) =\frac{1}{L_x L_y} \sum_{{\bf r}, {\bf r}'} e^{i {\bf k} \cdot ({\bf r} -{\bf r}')} \langle { s}^+_{\bf r} {s}^-_{{\bf r}'}  \rangle \; ,
\end{eqnarray}
evaluated at the AFM wave vector ${\bf k}=(\pi,\pi)$  (inset of Fig.~\ref{fig:4}).  $S^{\pm}(\pi,\pi)$ tends to a  value of order one for $L_x,L_y \to \infty$, indicating that  the  the order parameter $\frac{1}{L_x L_y} \sum_{{\bf r}} e^{i {\bf k} \cdot {\bf r} } {s}^{\nu}_{\bf r} $,  $\nu=\{x,y\}$, is proportional to $1/\sqrt{L_x L_y}$.

\begin{figure}[ht]
  \centering
  \includegraphics[width=8cm]{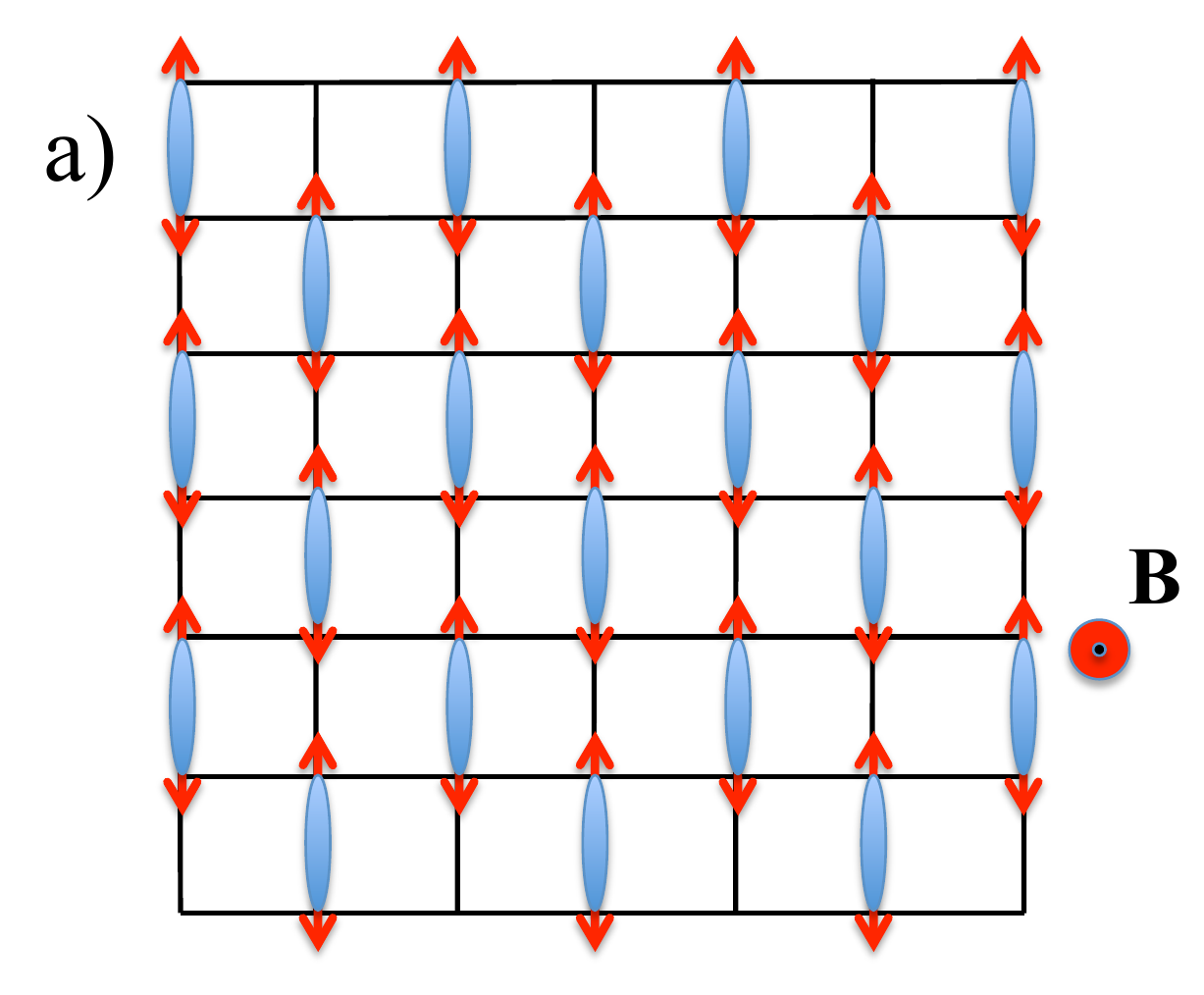}
  \caption{ Illustration of one of the possible scenarios in presence of a finite magnetic field ${\bf B}$. The ovals indicate that the corresponding bonds have a predominant singlet character. The staggered bond ordering (broken Z$_4$ symmetry) shown in the figure should produce a sharp maximum in the bond structure factor $S_B ({\bf k},\nu)$ at ${\bf k}=(\pi,\pi)$. The arrows indicate the AFM ordering 
in the plane perpendicular to the applied magnetic field induced by the condensation of triplons. The staggered magnetization can point along any direction obtained by a global spin rotation along the field direction (broken U(1) symmetry).}
  \label{fig:bos}
\end{figure}

This incipient AFM ordering occurs at the same point where an exponentially large number of different valence orderings become degenerate ($S=0$ ground state sector). However,  the presence of a triplon that propagates across the lattice could lead to two different scenarios for the bond correlations. It can either select one particular valence bond ordering via an order by disorder mechanism, or simply
destroy any long range bond ordering. In the former scenario, the application of a small magnetic field, $B$, that couples to the spins via the Zeeman term $- B \sum_{\bf r} S^z_{\bf r} $, should stabilize a particular bond ordering out of the exponentially large number of degenerate bond ordered ground states that exist at $B=0$. The selection mechanism would be provided by the kinetic energy of the field induced triplons, which should be minimized for a particular bond ordered background. In this scenario, the selected bond ordering coexists with the AFM $xy$-ordering induced by the condensation of triplons at the single particle state with momentum ${\bf k}=(\pi,\pi)$.
Figure~\ref{fig:bos} illustrates this situation for the bond ordering that has the highest susceptibility, according to the results that we discuss below. In the second scenario, the bond fluctuations induced by the triplon propagation are strong enough to produce a valence bond liquid with short-range bond-bond correlations. The only order parameter that survives is the AFM ordering which arises from the triplon condensation.

\begin{figure}[htp]
  \centering
  \includegraphics[width=9cm]{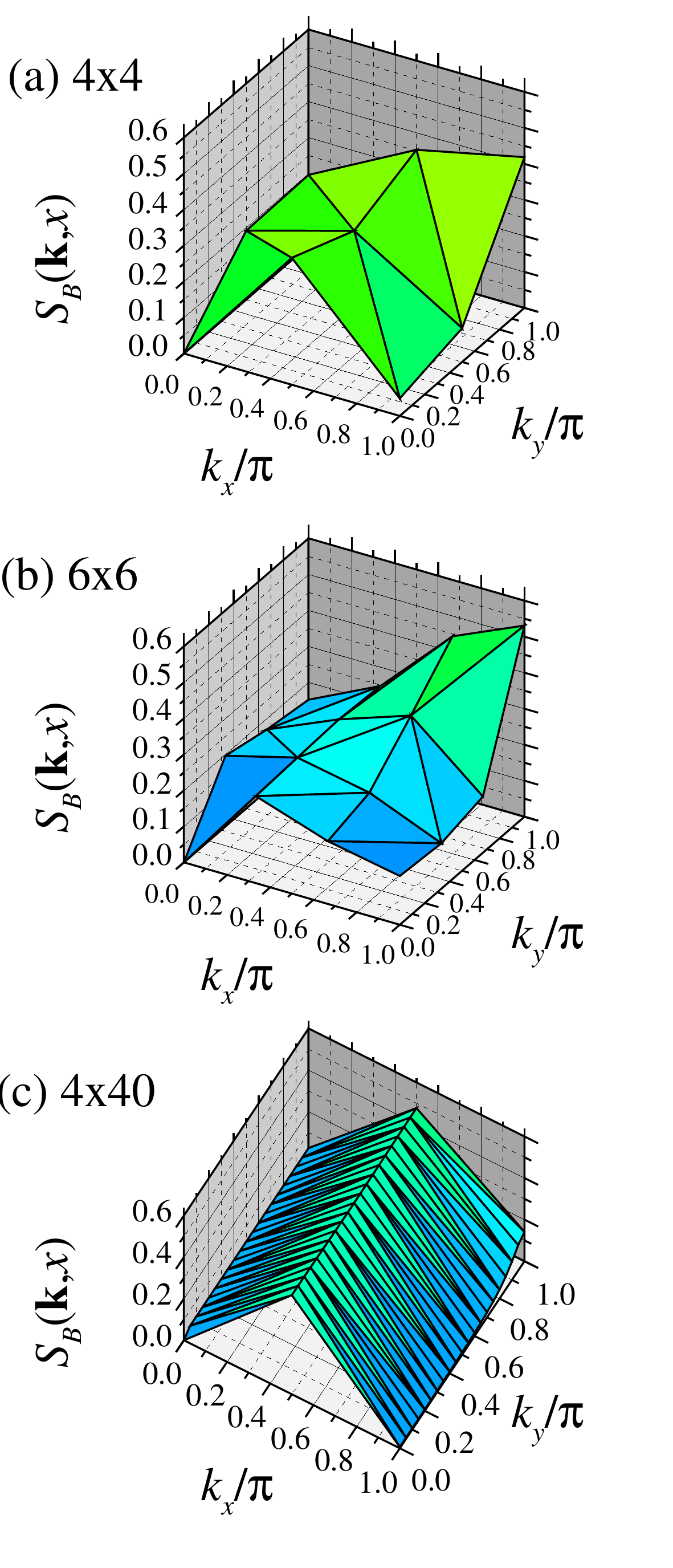}
  \caption{ Bond structure factors $S_B ({\bf k},x)$ as a function of ${\bf k}=(k_x,k_y)$, computed with the ERM for different lattice sizes. a) $L_x = L_y = 4 $. b) $L_x=L_y=6$. c) $L_x=4$, $L_y= 40$. }
  \label{fig:sqdimer}
\end{figure}

To further explore both scenarios, it is necessary to compute bond-bond correlation functions for the $S= S^z=1$ ground state. We introduce the bond structure factor, $S_B ({\bf k},\nu) $, for the local bond operator $B_{{\bf r}\nu} = {\bf s}_{\bf r} \cdot {\bf s}_{{\bf r}+{\bf e}_{\nu}} $. ${\bf k} = (k_x,k_y)$ is the wave vector, $\nu=\{ x,y \}$, and ${\bf e}_{\nu}$ is the relative vector connecting nearest-neighbor spins along the $\nu$-direction. Then, 
\begin{equation}
S_B ({\bf k},\nu) = \frac{1}{L_x L_y} \sum_{{\bf r}, {\bf r}'} e^{i {\bf k} \cdot ({\bf r} -{\bf r}')} \langle B_{{\bf r}\nu} B_{{\bf r}'\nu} \rangle \; .
\label{bbc}
\end{equation}
The staggered bond ordering (SBO) shown in Fig.~\ref{fig:bos} should produce a sharp maximum in $S_B ({\bf k},\nu)$ at ${\bf k}=(\pi,\pi)$.
In contrast, the bond structure factors obtained for $4\times4$, $6 \times 6$  lattices have very broad maxima at ${\bf k}=(\pi,\pi)$, indicating that the second scenario with short ranged bond correlations is more appropriate  for the $(S=1, S^z=1)$ ground state
(see Fig.~\ref{fig:sqdimer}~a and b). The same scenario hods for the $4 \times 40$ lattice (Fig.~\ref{fig:sqdimer}~c). In this case, $S_B({\bf k},y)$ has an approximately degenerate line of maxima for $k_x=\pi/2$, indicating that bond correlations between adjacent vertical lines are very weak.

\section{Application to a $D=1$ anisotropic Heisenberg model with NNN interactions}
\label{sec:1DHeisenberg}


The  one-dimensional family of
spin-1/2  Hamiltonians introduced in Refs.~\cite{Batista09} and \cite{Batista12} is
\begin{eqnarray}
\nonumber
H_{ Q} =\!\!\!\sum_{j;\nu=1,2} J_{\nu} \left [ \Delta_\nu \left ( s^z_{j+\nu} s^z_j - \frac{1}{4} \right)
+ s^x_{j+\nu} s^x_{j} + s^y_{j+\nu} s^y_{j}  \right].
\label{eq:Ham}
\end{eqnarray}
The Hamiltonian coefficients are parametrized   by the single variable $Q$: 
$\Delta_\nu = \cos (\nu Q)$ and $J_1 = -4J_2 \cos Q$, with $0 \le Q <\pi$.
 $H_{ Q}$ satisfies the FF property
for all $Q$ if we choose appropriate boundary conditions. In particular, for $Q= \pi/2$, the model reduces to two decoupled ferromagnetic Heisenberg chains whose
exact solutions are known~\cite{20.}. In this case, the ground-space dimension is $g=(L/2+1)^2$ or $g=(L+3)(L+1)/4$ for even or odd $L$, respectively. $L$ is the number of spins in the chain. 
The analytical solutions for the ground states  presented in Ref.~\cite{Batista12} show that the ground space
for $Q= \pi/2$ is continuously connected with the ground space for arbitrary $Q$. While, in principle,
the ground-space degeneracy for arbitrary $Q$ could be larger than that for $Q=\pi/2$, the ERM
shows that such a degeneracy remains constant.  Therefore, the ERM allows for an efficient computation of spin-spin correlations  in any ground state of $H_{ Q}$.




\section{Conclusions}
\label{sec:conclusions}

We  introduced an exact renormalization method that obtains the full ground-space of Hamiltonians
satisfying a frustration-free property.
The method outputs a sequence of tensors whose contraction allows for the computation
of correlation functions in any ground state. Such correlations can be used 
to characterize zero-temperature  states of matter.
The cost of the method depends on the ground state degeneracy 
for each renormalization step. The method computes the degeneracy and verifies the frustration-free
property.


We applied the method to two spin systems. First, we considered a FF Hamiltonian in $D=2$
whose ground-space dimension increases exponentially in the linear size of a square lattice. 
We applied the ERM successfully  and characterized the physical properties
of the only  ground state with total spin $S=1$ and projection $S^z=1$. Such a ground state differs qualitatively from the (bond ordered) singlet ground states identified in Ref.~\cite{Batista04}. In particular, it describes a quantum critical point associated with the onset of AFM ordering. According to our results, the application of an arbitrary small magnetic field $B>0$ induces AFM order of the spin components that are orthogonal to field's direction. Our results also indicate that the finite magnetic field destroys any of the long-range bond orderings that compete at $B=0$~\cite{Batista04}. 

We  also applied the ERM to the  one-dimensional family of FF spin-1/2 Hamiltonians described in Eq.~\eqref{eq:Ham}.
We verified that the ground space dimension is  $g = (L/2+1)^2$ ($g= (L+3)(L+1)/4$) for even (odd) $L$, as conjectured in Ref.~\cite{Batista12}. The ERM is efficient in this case.

Both applications illustrate the power of the ERM for obtaining exact ground state properties of frustration free Hamiltonians. 
Nevertheless, our renormalization method can also be used in more general contexts.
For example, since stochastic matrices can be sometimes related to FF Hamiltonians~\cite{stochastic}, the ERM
can be used to compute properties of (classical) systems in and out of equilibrium. Similarly, the ERM
can be used to solve combinatorial optimization problems~\cite{Somma07}.


\section{Acknowledgements}

Work at LANL was performed under the auspices of the U.S.\ DOE contract No.~DE-AC52-06NA25396 through the LDRD program. AEF thanks NSF for funding under grant No. DMR-0955707.
RDS thanks NSF for funding under the CCF program.


\begin{appendix}
\section{Exactness and cost of the ERM}
\label{app:RM} 
We provide more details about the ERM, showing the exactness of the method and 
estimating the computational requirements.
With no loss of generality $\pi_{v_k} $ is  
\begin{align}
\label{eq:pi-decomp}
\pi_{v_k} =\sum_{\alpha=1}^{A_k} O_{k,\alpha}^{w_{k(1)}} \otimes \ldots \otimes O_{k,\alpha}^{w_{k(r)}} \; ,
\end{align}
where $\otimes$ is the tensor product. The sets $w_{k(j)}$ satisfy $w_{k(j)} \cap v_k \ne \{ \emptyset \}$.
Any state $\ket{i^{v_k}}$ corresponds to a particular state $\ket{i^{w_{k(1)}}, \ldots ,i^{w_{k(r)}}}$ for the 
sets $w_{k(j)}$ in $v_k$. We assume then a
 specification of $H$  via access to $\Pi_H(.)$
that, on input $( v_k , i, i' )$, it outputs all the 
matrix elements
\begin{align}
\label{eq:Hentry}
\bra{i'^{w_{k(i)}}} O_{k,\alpha}^{w_{k(j)}} \ket{i^{w_{k(i)}}} \; ,
\end{align}
for $1 \le j \le r$ and  $1 \le \alpha A_k$. $\Pi_H$ also outputs all the sets $w_{k(j)}$ involved
in each term of the decomposition of $\pi_{v_k}$. In other words, $\Pi_H$ gives all the 
information about the action of each term of $\pi_{v_k}$ in the states $\ket{i^{v_k}}$.
Such a Hamiltonian specification is common in applications, e.g. a spin-1/2 system
 specified in terms of Pauli operators. The actual computational cost of the ERM
 should consider the number of times that  $\Pi_H$ is used, which is typically linear in $p$.

Our renormalization method
takes the ordered set $\{v_1, \ldots v_p\}$ as input  and uses
 $\Pi_H$. It outputs a bit $b$ denoting whether 
 $H$ is frustration free ($b=0$) or not ($b=1$)~\cite{NOTE3}.
 If $b=0$, the ERM also outputs the ground states
 specified by a sequence of tensors $T_l$ or $T^\dagger_l$ --
 see Eq.~\eqref{eq:groundstates}.
 Because Eq.~\eqref{eq:groundstates} involves a summation over repeated
 indexes, we use Einstein notation in the following.
For $l=1,\ldots,p$, $T_l$ is determined by its (complex) entries:  
 \begin{align}
T_{l}: = \left \{ \begin{matrix}   (t_1)_{i^{w_1}} ^{i^{z_1}} & {\rm if \ } l=1 \cr
& \cr
 (t_{l})_{i^{z_{l-1}} , i^{w_l}} ^ {i^{z_l}} & {\rm if \ } l>1
 \end{matrix}
  \right. \; ,
\end{align}
with $1 \le i^{w_l} \le d_{w_l}$, $1 \le i^{z_l} \le g_l$,  and $g_l$ determined by the ERM
(see below).
For each $i^{z_l} \le g_l$,
the  vectors $\ket{\psi^l_{i^{z_l}}}=T_1^\dagger \bullet ... \bullet T_{l}^\dagger  \ket{ i^{z_{l}}}$ 
are in $\cH_{z_l}$ and thus have $d_{z_l}$ components in the computational basis. 
We recall that $z_l = \cup_{k=1}^l w_k$. Then, each such
component is determined
by  $i^{w_1}, \ldots, i^{w_l}$ corresponding to the 
sets $w_1, \ldots , w_l$, respectively. That is, $\{ \ket{i^{w_1},\ldots ,i^{w_l}} \}$,
with $1 \le i^{w_k} \le d_{w_k}$,
also defines a computational basis for $\cH_{z_l}$.  
The components of $\ket{\psi^l_{i^{z_l}}}$ in such a basis are
  \begin{align}
  \label{eq:treetensor}
&    (u_1)^{i^{w_1}} _{i^{z_1}} (u_2)^{i^{z_1},i^{w_2}} _{i^{z_2}}  \ldots  (u_{l})^{i^{z_{l-1}} , i^{w_l}} _ {i^{z_l}}  \; ,
\end{align}
 with 
 \begin{align}
 (u_1)^{i^{w_1}} _{i^{z_1}} &= ((t_1)_{i^{w_1}} ^{i^{z_1}})^* \\
 \nonumber
 (u_{k})^{i^{z_{k-1}} , i^{w_k}} _ {i^{z_k}} & = ((t_{k})_{i^{z_{k-1}} , i^{w_k}} ^ {i^{z_k}}  )^*  \; , \; 2 \le k \le p \;.
 \end{align}
  In particular, when $l=p$, we have $z_p = \cV$ and Eq.~\eqref{eq:treetensor}
  defines the contraction in Eq.~\eqref{eq:groundstates}.
 This contraction is associated with a tree-like tensor
 network -- see Fig~\ref{fig:RGFig2} for an example.

  For $l=1$, we denote by $\gamma_1$  the $d_{v_1}$-dimensional matrix representation of $\pi_{v_1}$ in $\cH_{v_1} = \cH_{w_1}$,
 in the computational basis. The ERM constructs $\gamma_1$ using $\Pi_H$ once. Then, the ERM
  performs {\em exact} diagonalization  
 to obtain $g_1$ and an orthonormal vector basis $\{ \ket{\phi^1_{i^{z_1}}} \}_{1 \le i^{z_1} \le g_1 }$
 for the zero-eigenvalue eigenvectors of $\gamma_1$.
 $T_1$ is the tensor that maps such eigenvectors
 to vectors or states in the computational basis: 
 \begin{align}
 T_1= \sum_{i^{z_1}=1}^{g_1} \ket{i^{z_1}} \bra{\phi^1_{i^{z_1}}} \; .
 \end{align}
 The entries of $T_1$ are
\begin{align}
 (t_1)_{ i^{w_1}} ^ {i^{z_1}} = \langle {\phi^1_{i^{z_1}}} \ket{i^{w_1}}  \; ,
\end{align}
with $1 \le i^{w_1} \le d_{w_1}$ and $1 \le z_1 \le g_1$. 

 In the $l$-th step, $l \ge 2$, the ERM uses $\Pi_H$ to construct the $h_l \times h_l$ matrix
 $\gamma_l$, $h_l=g_{l-1}.d_{w_l}$, with entries
 \begin{align}
 \label{eq:gammal}
 \gamma_l \rightarrow \bra{\phi^{l-1}_{i'^{z_{l-1}}}, i'^{w_l}} \pi_{v_l}  \ket{\phi^{l-1}_{i^{z_{l-1}}}, i^{w_l}} \; .
 \end{align}
Here, $ i^{z_{l-1}}, i'^{z_{l-1}} = 1,\ldots,g_{l-1}$  
and  $i'^{w_l} , i^{w_l}=1,\ldots,  d_{w_l}$.
The ERM performs exact diagonalization of $\gamma_l$ and continues
 only if the lowest eigenvalue is 0 ($b=0$).
It computes the multiplicity of the zero eigenvalue, assigns
 it to $g_l$, and computes a complete orthonormal basis of 
 $h_l$-dimensional eigenvectors  $\{ \ket{\phi^l_1} , \ldots , \ket{\phi^l_{g_l}} \}$
 for the zero eigenvalue. 
The ERM  assigns the tensor $T^{\;}_l$ to the transformation that maps such eigenvectors
to vectors in the computational basis of $\cH_{z_l}$:
\begin{align}
\label{Isometry2}
T_l^{\;} = \sum_{i^{z_l}= 1}^{g_l} \ket {i^{z_l}} \bra{\phi^l_{i^{z_l}}} \; .
\end{align}
The entries of $T_l$ are 
\begin{align}
 (t_l)_{i^{z_{l-1}} , i^{w_l}} ^ {i^{z_l}} = \langle \phi^l_{i^{z_l}}  \ket{i^{z_{l-1}} , i^{w_l}} \; .
\end{align}

To show that the ERM is exact, we first remark that
the eigenvectors $\ket{\phi^l_{i^{z_l}}}$ represent
 the states $\ket{\psi^l_{i^{z_l}}}$, as determined by Eq.~\eqref{eq:treetensor}.
Then, we will show
that the states
\begin{align}
\ket{\psi^l_{i^{z_l}}} = T_1^\dagger \bullet \ldots \bullet T^\dagger_l \ket{i_z^l} \; ,
\end{align}
given according to Eq.~\eqref{eq:treetensor}, are ground states of $H_l=\sum_{k=1}^l \pi_{v_k}$ for all $l \in \{1,2,\ldots, p\}$,
when $H_l$ are frustration free.  

The proof is inductive. For $l=1$, 
$\ket{\phi^1_{i^{z_1}}} = \ket{\psi^1_{i^{z_1}}}$
is a ground state of $\gamma_1$ or $\pi_{v_1}=H_1$ by definition. 
 That is,  $\ket{\psi^1_{i^{z_1}}}$, as determined from Eq.~\eqref{eq:treetensor}, is a ground state of $H_1$ for each $i^{z_1}=1, \ldots, g_1$.
We assume now that 
\begin{align}
\nonumber
\{ \ket{\psi^{l-1}_{i^{z_{l-1}}}} = T_1^\dagger \bullet \ldots \bullet T^\dagger_{l-1} \ket{i_z^{l-1}} \}_{1 \le i^{z_{l-1}}
\le g_{l-1}}
\end{align}
is an orthogonal basis for the ground subspace of $H_{l-1}$.
Then, if $\ket \phi$ is a ground state of $H_l$,
\begin{align}
\label{eq:generalGS}
\ket{\phi} = \sum_{i^{z_{l-1}}=1}^{g_{l-1}} \sum_{i^{w_l}=1}^{d_{w_l}} c_{i^{z_{l-1}},i^{w_l}} \ket{\psi^{l-1}_{i^{z_{l-1}}} , i^{w_l}}
\end{align}
with $c_{i^{z_{l-1}},i^{w_l}}$ complex amplitudes. Otherwise,
$\ket{\phi}$ would not belong to the intersection between the ground subspaces of $H_{l-1}$
and $H_l$, a requirement for frustration-free Hamiltonians.
It   suffices to obtain $\gamma_l$, a projection of $\pi_{v_l}$ into the subspace spanned by
$\{ \ket{\psi^{l-1}_{i^{z_{l-1}}} , i^{w_l}} \} _{1 \le i^{z_{l-1}} \le g_{l-1} , 1 \le i^{w_l} \le d_{w_l} }$.
Without loss of generality, we can choose a value of $i^{z_l}$ such that
\begin{align}
\ket \phi =\ket{\phi^l_{i^{z_{l}}}}  \; .
\end{align}
Thus,
\begin{align}
 c_{i^{z_{l-1}},i^{w_l}} = (u_l)_{i^{z_{l-1}},i^{w_l}}^{i^{z_l}} \; ,
\end{align}
where $(u_l)^{i^{z_{l-1}},i^{w_l}}_{i^{z_l}}$ are the entries of $T_l^\dagger$.
That is,
\begin{align}
\label{eq:Hl-groundstate}
& \ket \phi  = (u_l)^{i^{z_{l-1}},i^{w_l}}_{i^{z_l}} \ket{\psi^{l-1}_{i^{z_{l-1}}} , i^{w_l}} \\
\nonumber
& =  (u_1)^{i^{w_1}}_{i^{z_1}} ( u_2)^{i^{z_{1}},i^{w_2}}_{i^{z_2}}   \ldots (u_l)^{i^{z_{l-1}},i^{w_l}}_{i^{z_l}} \ket{i^{w_1}, \ldots , i^{w_l}} \; .
\end{align}
The contraction in Eq.~\eqref{eq:Hl-groundstate} coincides with that of Eq.~\eqref{eq:treetensor}.
It follows that
\begin{align}
\ket \phi = \ket{\psi^l_{i^{z_{l}}}} = T^\dagger_1 \bullet \ldots \bullet T^\dagger_l \ket{i^{z_l}}
\end{align}
is a ground state of $H_l$ for all $1 \le i^{z_l} \le g_l$. In particular, $ \ket{\psi^p_{i^{z_{p}}}}$,
with $1 \le i^{z_p} \le g_p=g$, are all the 
ground states of $H$.

\subsection{Computational requirements}
\label{sec:cost}
 We let $N_T$ be the total cost, i.e., the total number
of elementary operations to obtain all entries of $T_1,
\ldots, T_p$. For simplicity, we do not consider in the cost the number of queries to
$\Pi_H$, which is typically linear in $p$. We also let $M_T$ be the memory
requirements, i.e., the number of coefficients that need to be kept in memory
during the implementation of he ERM.

We first write
\begin{equation}
N_T = \sum_{l=1}^p N_T^l \; , \; M_T = \sum_{l=1}^p M_T^l \; .
\end{equation}
To obtain $T_1$, the ERM performs exact diagonalization of $\gamma_1$.
Because
$\gamma_1$ is of dimension $d_{w_1}$, the 
cost of obtaining its eigenvectors and eigenvalues is   $N_T^1 \le EV(d_{w_1})
\in \cO({\rm poly}(d_{w_1}))$.  $EV(d)$ is the cost of exact diagonalization
of  a $d \times d$ matrix, which is almost quadratic
in $d$ in actual implementations.  Only the $g_1$  eigenvectors 
with zero eigenvalue need 
to be kept in memory for the following step and thus $M_T^1 \propto g_1 d_{w_1}$.

\begin{figure}[ht]
  \centering
  \includegraphics[width=6.cm]{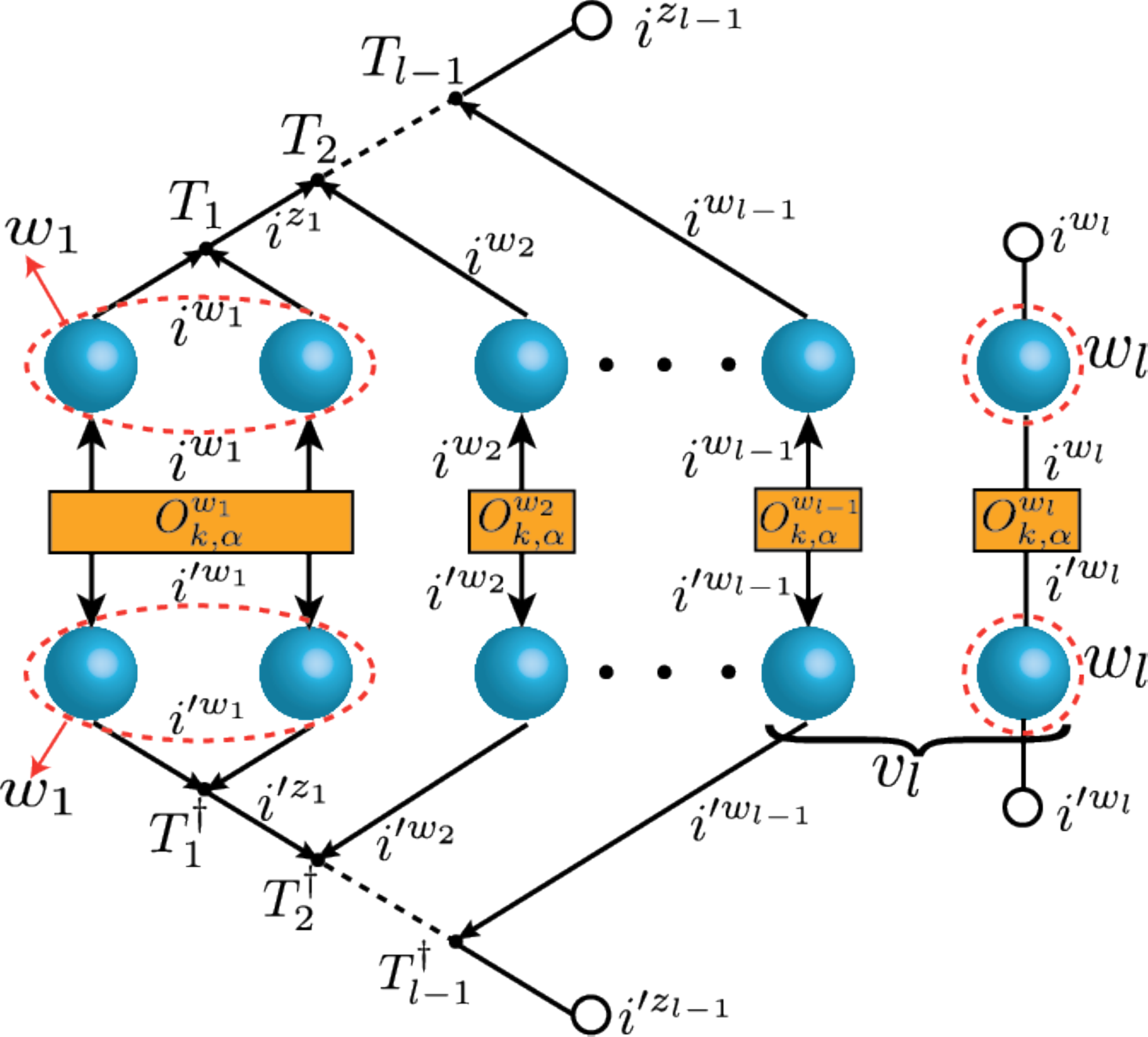}
  \caption{Representation of the network contraction
  for each matrix element of $\gamma_l$ in Eq.~\eqref{eq:tensorgamma3}
  for the same system of Fig.~\ref{fig:RGFig2}. Blue circles are spins.   $\pi_{v_l}$
  is a sum of $A_k$ terms of the form $O_{k,\alpha}^{w_1} \otimes \ldots
  \otimes O_{k,\alpha}^{w_l}$, and some $O_{k,\alpha}^{w_j}$ may act
  trivially in $w_j$; i.e.,  $O_{k,\alpha}^{w_j}= \sum_{i^{w_j}=1}^{d_{w_j}} \ket{i^{w_j}}\bra{i^{w_j}}$.
  Arrows denote a sum of the  index in the corresponding edge.
  Open circles denote a fixed index, referring to a particular matrix element of $\gamma_l$,
  the projection of $\pi_{v_l}$ in the ground subspace of $H_{l-1}=\sum_{k=1}^{l-1} \pi_{v_k}$.
  $1 \le i^{w_l},i'^{w_l} \le d_{w_l}$ and
  $1 \le i^{z_{l-1}},i'^{z_{l-1}} \le g_{l-1}$. }
  \label{fig:RGFig3}
\end{figure}

The cost
of obtaining all zero-eigenvalue eigenvectors of $\gamma_l$ is bounded by
$EV(h_l)$.  To obtain $N_T^l$
we need to add the cost of computing $\gamma_l$.
Each matrix element of $\gamma_l$ in Eq.~\eqref{eq:gammal}
is
\begin{align}
\label{eq:tensorgamma}
 \bra{i'^{z_{l-1}}} \bra { i'^{w_l}} T^{\;}_{l-1} \bullet \ldots \bullet T^{\;}_1 
  \bullet \pi_{v_l} \bullet \\
  \nonumber \bullet  T^\dagger_1 \bullet \ldots \bullet T^\dagger_{l-1} \ket{i^{z_{l-1}}} \ket { i^{w_l}} \; .
\end{align}
We consider the decomposition in Eq.~\eqref{eq:pi-decomp} and 
we are interested in obtaining the cost of computing a particular term
\begin{align}
\label{eq:tensorgamma2}
 \bra{i'^{z_{l-1}}} \bra { i'^{w_l}} T^{\;}_{l-1} \bullet \ldots \bullet T^{\;}_1 
  \bullet \left(O_{k,\alpha}^{w_{k(1)}} \otimes \ldots \right.
  \\
  \nonumber
  \left. \ldots \otimes O_{k,\alpha}^{w_{k(r)}} \right) \bullet  T^\dagger_1 \bullet \ldots \bullet T^\dagger_{l-1} \ket{i^{z_{l-1}}} \ket { i^{w_l}} \; .
\end{align}
We can interleave trivial operators $\one_w=\sum_{i^w=1}^{d_w} \ket{i^w}\bra{i^w}$ in Eq.~\eqref{eq:tensorgamma2}
for those $w \ne w_{k(j)}$ without affecting the output of the contraction.
That is, we extend the definition of $O_{k,\alpha}^{w_{k(j)}}$ so that
$O_{k,\alpha}^{w_{k(j)}} = \one_{w_{k(j)}}$ for $r < j \le l$.
Then we write
\begin{align}
(o_j)_{i'^{w_{k(j)}}}^{i^{w_{k(j)}}}=\bra{i'^{w_{k(j)}}} O_{k,\alpha}^{w_{k(j)}} \ket{i^{w_{k(j)}}} \; 
\end{align}
for all $1 \le j \le l$.
Equation~\eqref{eq:tensorgamma2} is 
\begin{align}
\label{eq:tensorgamma3}
(t_1)_{i'^{w_1}} ^{i'^{z_1}}    \ldots  (t_{l-1})_{i'^{z_{l-2}} , i'^{w_{l-1}}} ^ {i'^{z_{l-1}}} \left[ (o_1)_{i'^{w_1}}^{i^{w_1}} \ldots \right.
\\
\nonumber
\left.
(o_l)_{i'^{w_l}}^{i^{w_l}} \right] (u_1)^{i^{w_1}} _{i^{z_1}}    \ldots  (u_{l-1})^{i^{z_{l-2}} , i^{w_{l-1}}} _ {i^{z_{l-1}}} \; ,
\end{align}
where we used Eq.~\eqref{eq:treetensor}.
As before, Eq.~\eqref{eq:tensorgamma3} refers to a contraction of a tree-like
tensor network; see Fig.~\ref{fig:RGFig3} for an example.

The cost of evaluating Eq.~\eqref{eq:tensorgamma3}
depends on the support of $v_l$, that is, the 
 number and position of spins that belong to $v_l$.
 This is so because Eq.~\eqref{eq:tensorgamma3}
 can be sometimes simplified considering that
 \begin{align}
 T^{\;}_k T^\dagger_k = \sum_{i^{z_k}=1}^{g_k} \ket{i^{z_k}} \bra{i^{z_k}}
 \end{align}
 and then
 \begin{align}
(t_k)^{i'^{z_k}}_{i^{z_{k-1}},i^{w_k}} (u_k)_{i^{z_k}}^{i^{z_{k-1}},i^{w_k}} = \delta_{i'^{z_k},i^{z_k}} \; ,
 \end{align}
 which are useful if $O_{k,\alpha}^{w_{k(j)}}=\one_{w_{k(j)}}$.
 Nevertheless, to analyze the cost of computing
 Eq.~\eqref{eq:tensorgamma3} we consider the worst case scenario in which
 $O_{k,\alpha}^{w_{k(j)}} \ne \one_{w_{k(j)}}$ for all $1 \le j \le l$.
 We can rearrange the sum and
 compute Eq.~\eqref{eq:tensorgamma3} in $l$ sequential steps as follows.
 First, we compute  the  
  \begin{align}
  \nonumber
(t_1)^{i'^{z_1}}_{i'^{w_1}}  (o_1)_{i'^{w_1}}^{i^{w_1}} (u_1)_{i^{z_1}}^{i^{w_1}}
 \end{align}
 for all $1 \le i^{z_1},i'^{z_1} \le g_1$. Because $1 \le i^{w_1},i'^{w_1} \le d_{w_1}$,
 this step has cost $\propto (d_{w_1} g_1)^2$. We keep the computed values in memory.
 Next we compute
 the
 \begin{align}
 \nonumber
 (t_1)^{i'^{z_1}}_{i'^{w_1}} (o_1)_{i'^{w_1}}^{i^{w_1}}
(u_1)_{i^{z_1}}^{i^{w_1}} (t_2)^{i'^{z_2}}_{i'^{z_1},i'^{w_2}} (o_2)_{i'^{w_2}}^{i^{w_2}}  (u_2)_{i^{z_2}}^{i^{z_1},i^{w_2}}
 \end{align}
 for all $1 \le i^{z_2},i'^{z_2} \le g_2$. This step has an additional cost
  $\propto (g_1 d_{w_2} g_2)^2$, where the first $g_1$ comes from the sum in
 $i^{z_1}$ and  $i'^{z_1}$.  We keep implementing the procedure sequentially
  until we compute
 \begin{align}
& (t_1)^{i'^{z_1}}_{i'^{w_1}}  (o_1)_{i'^{w_1}}^{i^{w_1}} (u_1)_{i^{z_1}}^{i^{w_1}} \ldots \\
 \nonumber
& \ldots (t_{l-1})^{i'^{z_{l-1}}}_{i'^{z_{l-2}},i'^{w_{l-1}}} (o_{l-1})_{i'^{w_{l-1}}}^{i^{w_{l-1}}} (u_{l-1})_{i^{z_{l-1}}}^{i^{z_{l-2}},i^{w_{l-1}}} \; ,
  \end{align}
 which has an additional cost  $\propto (h_{l-1} g_{l-1})^2$ with respect to the previous computations.
 That is,  the sequential method has an overall cost  $\propto \sum_{k=1}^{l-1} (h_k.g_k)^2$,
 with $g_0=1$. The sequential method can be understood from the
 example in Fig.~\ref{fig:RGFig3}, in which the sequential steps regard the contraction of tensors
 from left to right.
 
 The last step  is to compute
 \begin{align}
 (o_l)_{i'^{w_l}}^{i^{w_l}}
 \end{align}
 for all $1 \le i^{w_l}, i'^{w_l} \le d_{w_l}$. This step is implemented using $\Pi_H$ and has no cost
 under our assumption.
 Then, the computation of Eq.~\eqref{eq:tensorgamma3} for all $1 \le i^{w_l}, i'^{w_l} \le d_{w_l}$
 and $1 \le i^{z_{l-1}}, i'^{z_{l-1}} \le g_{l-1}$ can be implemented with
 \begin{align}
 \propto \sum_{k=1}^{l-1} (g_{k-1} d_{w_k}.g_k)^2
 \end{align}
 elementary operations. To compute  $\gamma_l$, we need to add a multiplicative factor $A_l$
 that regards the number of terms in the decomposition of $\pi_{v_l}$.
 Then,
 \begin{align}
 N_T^l =   EV(g_{l-1} d_{w_l})  + c A_l \sum_{k=1}^{l-1} (h_k .g_k)^2 \; ,
 \end{align}
 with $c >1$ a constant. 
 A memory of $M_T^l \propto h_l g_l$ is needed for
 the ground states of $\gamma_l$.

 Typically, $A_l$ is bounded by some constant $A$.
 In this case,
 \begin{align}
 \nonumber
 N_T & \propto \sum_{k=1}^p EV(h_k)  + (p-1) (d_{w_1} g_1)^2 + \\
 \label{eq:totalcost}
 &+  (p-2)(h_2 g_2)^2+ \ldots + (h_{p-1} g_{p-1})^2 \; .
 \end{align}
 In addition,
 \begin{align}
 M_T \propto \sum_{k=1}^p h_k g_k \; .
 \end{align}
If $g_k \in \cO[{\rm poly} (p)]$ and $d_{w_k}  \in \cO[{\rm poly} (p)]$,
 then $N_T \in  \cO[{\rm poly} (p)]$ and the ERM is efficient. 
 
\subsection{Optimal cost}
The total cost of the ERM  in Eq.~\eqref{eq:totalcost} depends on 
$d_{w_l}$ and $g_l$. In many applications, $d_{w_l}$ is constant and 
$N_T$ and $M_T$ are functions of $g_1, \ldots ,g_p$. The cost can then be minimized
by considering all possible orderings of $w_1, \ldots, w_p$ such that $N_T$ and/or $M_T$ are minimum.
This procedure rules out some possible orderings that yield exponential complexity 
in systems in which the ERM could be implemented efficiently. For example,
consider a square spin lattice and assume that the terms $\pi_{v_k}$ in the Hamiltonians involve two nearest-neighbor spins
in either direction. The  ERM  could be implemented to
obtain the ground states of each chain along a particular direction
and then add the Hamiltonian terms in the other direction. However, this construction
results in exponential complexity if the ground space of each chain is degenerate because the
number of ground states that have to be kept in memory is exponentially large in the  length
of the chains. A more efficient choice considers
  ``growing'' the system using thr ``snake'' path depicted in Fig.~\ref{fig:lattice} (b).

In a different example we consider a binary tree of depth $q$ and assume $p=2^q$.
Each node in the basis of the tree corresponds to a single spin in a lattice.
A standard real-space renormalization method for such a binary tree will have $l=1,\ldots,q$ steps, each involving a diagonalization
of $2^{q-l}$ matrices of dimension $g_{2^l}$ each.  The
memory requirement for such method is dominated by the last step, which 
requires dealing with a subspace of dimension $g_{p/2} \times g_{p/2}$,
spanned by all the ground states obtained in the previous step. In addition, each such
ground state has $(g_{p/4})^2$ components in a computational basis. If 
 $g_k \propto k^{\beta}$, the memory requirement to implement the last step
 is   $M'_T \in \cO[p^{4 \beta}]$. Nevertheless, our ERM implies $M_T \in \cO(p^{2\beta})$ in this case.
 Clearly, $M'_T \gg M_T$ when $\beta >0$, $p \gg 1$. 
 The standard renormalization method may outperform the ERM only if $\beta=0$.
 
\end{appendix}



\begin{thebibliography}{10}
 
\bibitem{Wilson75} 
K.G. Wilson, Rev. Mod. Phys. 47, 773 (1975).

\bibitem{BullaRMP08}
R. Bulla, T.~A. Costi and T. Pruschke, Rev. Mod. Phys. {\bf 80}, 395 (2008).

\bibitem{Kondo64}
J. Kondo,  Prog. Theor. Phys. {\bf 32}, 37 (1964). 
 
\bibitem{White92}
S.R. White, Phys. Rev. Lett. 69, 2863 (1992).

\bibitem{steve kagome} Simeng Yan, David A. Huse, and Steven R. White, Science {\bf 332} 1173 (2011).

\bibitem{peps matthias} P. Corboz, S. R. White, G. Vidal, M. Troyer, Phys. Rev. B {\bf 84}, 041108 (2011).

\bibitem{uli kagome} S. Depenbrock, I. P. McCulloch, U. Schollw\"ock, Phys. Rev. Lett. {\bf 109}, 067201 (2012).

\bibitem{balents j1j2}  H.-C. Jiang, H. Yao, L. Balents, Phys. Rev. B {\bf 86}, 024424 (2012).

\bibitem{peps2d} F. Verstraete and J. I. Cirac, cond-mat/0407066. V. Murg, F. Verstraete, and J. I. Cirac, Phys. Rev. A {\bf 75}, 033605 (2007).


\bibitem{Sab08}
Hamed Saberi, Andreas Weichselbaum, and Jan von Delft, Phys. Rev. B {\bf 78}, 035124 (2008).




\bibitem{mera} Guifre Vidal, Phys. Rev. Lett.{\bf 99}, 220405 (2007), G. Evenbly, G. Vidal, Phys. Rev. B {\bf 79}, 144108 (2009).

\bibitem{mera2d} Glen Evenbly, Guifre Vidal, Phys. Rev. Lett. {\bf 102}, 180406 (2009).

\bibitem{trg} G. Sierra and M. A. Martn-Delgado, cond-mat/9811170
(preprint); Y. Nishio, N. Maeshima, A. Gendiar, and T.
Nishino, cond-mat/0401115.

\bibitem{levin} Zheng-Cheng Gu, Michael Levin, Xiao-Gang Wen, Phys. Rev. B {\bf 78}, 205116 (2008).

\bibitem{tao} Z. Y. Xie, H. C. Jiang, Q. N. Chen, Z. Y. Weng, T. Xiang, Phys.Rev.Lett. {\bf 103} 160601 (2009).

\bibitem{tao2} H.H. Zhao, Z.Y. Xie, Q.N. Chen, Z.C. Wei, J.W. Cai, T. Xiang, Physical Review B {\bf 81}, 174411 (2010).



\bibitem{balents} H.-C. Jiang, Z. Wang, L. Balents, Nature Phys. {\bf 8}, 902-905 (2012).


\bibitem{Perez08}
D. Perez-Garcia, F. Verstraete, M.M. Wolf, and J.I. Cirac, Quantum
Information and Computation {\bf 8}, 650 (2008).

\bibitem{Bravyi2009}
S. Bravyi and B. Terhal, SIAM J. Comp. {\bf 39}, 1462 (2009)

\bibitem{Somma2011}
R.D. Somma and S. Boixo, e-print arXiv:1110.2494 (2011).

\bibitem{AKLT}
I. Affleck, T. Kennedy, E.H. Lieb, and H. Tasaki, 
Phys. Rev. Lett. {\bf 59}, 799 (1987).

\bibitem{Aharonov2004}
D. Aharonov, W. Van Dam, J. Kempe, Z. Landau, S. Lloyd, and O. Regev,
in {\em Proc. of the 45th Annual IEEE Symp. on Found. Comp. Sci.},
42 (2004).


\bibitem{Batista12}
C. D. Batista and R. D. Somma, Phys. Rev. Lett. 109, 227203 (2012).


\bibitem{Wannier50}
G. H. Wannier,  Phys. Rev. {\bf 79}, 357 (1950).

\bibitem{Batista04}
C. D. Batista and S. Trugman, Phys. Rev. Lett. {\bf 93}, 217202 (2004).

 
\bibitem{Lajko12}
M. Lajk\'{o}, P. Sindzingre, and K. Penc, Phys. Rev. Lett. 108, 017205 (2012).
 
 \bibitem{Bravyi2011}
 S. Bravyi.  Contemporary Mathematics {\bf 536}, 2011.
 
 \bibitem{BOO2010}
 N. de Beaudrap, M. Ohliger, T.J. Osborne, and J. Eisert,
 Phys. Rev. Lett. {\bf 105}, 060504 (2010).


\bibitem{HS10}
Hong-Hao and Mikel Sanz, Phys. Rev. B {\bf 82}, 104404 (2010).

\bibitem{NOTE2}
The ERM also works if $\cal V$ corresponds to a set of sites
that can be occupied by a limited number of particles like fermions or bosons~\cite{JW1928}.

\bibitem{JW1928}
P. Jordan and E. P. Wigner, Z. Phys. 47, 631 (1928).


\bibitem{NOTE1}
Note that, by using the superscript $x$, we also clarify
the Hilbert space to which $\ket{i^x}$ belongs to.

\bibitem{Vid2007}
G. Vidal, Phys. Rev. Lett. {\bf 99}, 220405 (2007).

\bibitem{EV2012}
G. Evenbly and G. Vidal, e-print arXiv:1205.0639 (2012).



\bibitem{note1}
Karlo Penc, private communication. 

\bibitem{Batista09}
C. D. Batista, Phys. Rev. B 80, 180406 (2009).


\bibitem{20.}
H. Bethe, Z. f\"ur Physik A {\bf 71}, 205 (1931).

\bibitem{stochastic}
C.L. Henley, J. of Phys: Cond.Mat. {\bf 16}, S891 (2004). C. Castelnovo,
C. Chamon, C. Murphy, and P. Pujol, Ann. of Phys. {\bf 318},
316 (2005). F. Verstraete, M. M. Wolf, D. Perez-Garcia, and J.
I. Cirac, Phys. Rev. Lett. {\bf 96}, 220601 (2006).

\bibitem{Somma07}
R.D. Somma, C.D. Batista, and G. Ortiz,
Phys. Rev. Lett. {\bf 99}, 030603 (2007).

\bibitem{NOTE3}
If $H$ is {\em almost} frustration free, 
our renormalization method may still provide
a good approximation to the ground states.
In this case, an upper bound to the error
can be computed by the method.



 \end{thebibliography}
\end{document}